\documentclass[aps,prx,twocolumn,letterpaper,superscriptaddress,showpacs,longbibliography]{revtex4-1}
\usepackage{graphicx}
\usepackage{CJK}
\usepackage{verbatim}
\usepackage{physics}
\usepackage[T1]{fontenc}
\usepackage[colorlinks,linkcolor=blue,anchorcolor=blue, citecolor=blue,urlcolor=blue,]{hyperref}
\usepackage{lineno} 
\usepackage{bm}
\usepackage{mathptmx}
\usepackage{needspace}
\usepackage[normalem]{ulem}
\makeatletter
\begin{document}
\begin{CJK*}{UTF8}{bsmi}
\title{Charge order through crystallization of Frenkel excitons: realization in kagome metals}
\author{Ruoshi Jiang}
\affiliation{School of Physics and Astronomy and Tsung-Dao Lee Institute, Shanghai Jiao Tong University, Shanghai 200240, China}
\affiliation{Department of Materials Science and Metallurgy, University of Cambridge, Cambridge CB3 0FS, United Kingdom}
\author{Bartomeu Monserrat}
\affiliation{Department of Materials Science and Metallurgy, University of Cambridge, Cambridge CB3 0FS, United Kingdom}
\author{Wei Ku}
\altaffiliation{corresponding email: weiku@sjtu.edu.cn}
\affiliation{School of Physics and Astronomy and Tsung-Dao Lee Institute, Shanghai Jiao Tong University, Shanghai 200240, China}
\affiliation{Key Laboratory of Artificial Structures and Quantum Control (Ministry of Education), Shanghai 200240, China}
\affiliation{Shanghai Branch, Hefei National Laboratory, Shanghai 201315, People's Republic of China}

\begin{abstract}
Charge order is a widely observed and representative example of spontaneous broken symmetries in quantum states of matter.
Owing to the large intra-atomic Coulomb energy, the charge redistribution in such an order typically implies significant alteration of the electronic and lattice properties of materials.
While the standard description of charge order, namely a ``charge density wave'' instability of the Fermi surface, has been broadly and successfully applied to good metals, its applicability to correlated ionic materials has been rather limited.
Here, we propose an alternative general scenario of charge order - crystallization of long-lived Frenkel excitons - suitable for these ionic materials.
We demonstrate this scenario on the recently discovered kagome superconductors and successfully reproduce all the characteristics of experimental observations on both local charge correlations and long-range ordering.
The proposed generic scenario offers a long-sought understanding of charge order applicable to modern correlated functional materials.
\end{abstract}
\maketitle
\end{CJK*}

\section*{Introduction}

The phenomenon of charge order (CO) has been widely observed in a variety of ionic materials (in which at least some atoms have a well-defined effective ionic valence), including manganates\,\cite{Mori1998, Chen1996Commensurate}, nickelates\,\cite{Staub2002, Junjie2016Stacked}, cobaltates\,\cite{Babkevich2016, Foo2014}, cuprates\,\cite{Keimer2015, Wen2019, Arpaia2021}, kagome metals\,\cite{Jiang2021Unconventional, Yin2022Discovery, Ma2025Correlation}, and transition metal dichalcogenides\,\cite{Arguello2014, Morosan2006Superconductivity, Wu2023Discovery}.
Due to the large energy scale of the Coulomb interaction, the charge redistribution and long-range ordering associated with the formation of CO are often found to strongly couple to observed physical properties such as transport\,\cite{Harper1975, Zhang2020Intertwined}, magnetism\,\cite{Hu2022,Chen2023}, and other symmetry-broken phases\,\cite{Ishkawa1999}.
Highlighting systems displaying superconductivity, besides the obvious competition in long-range ordering\,\cite{Titus2022, Eduardo2014}, the relationship between short-range charge and superconducting correlations remains an open question that attracts intensive research activities\,\cite{Hayden2024Charge, Porter2024Understanding}.

The standard description of charge order, namely the theory of charge density waves (CDWs), is based on interaction-induced Fermi surface instabilities\,\cite{Peierls1930, Peierls1955quantum,Frohlich1954On, Aebi2001, Kohn1959Image, Rice1975New,McMillan1977Microscopic,Varma1983Strong, Johannes2008Fermi, Xu2021Topical, Rossnagel2011On}, and it naturally applies to good metals in which the kinetic energy dominates the electronic structure.
In semimetals with low carrier densities and in semiconductors with small gaps, these instabilities can be further formulated through the formation of Wannier excitons\,\cite{Kohn1967,Halperin1968,Dove1993,Monique2015Excitons,Anshul2017Signatures}, which can even drive the system into an excitonic insulator phase\,\cite{Kohn1967,Jerome1967Excitonic} under strong enough electron-hole binding energies.
Nonetheless, applicability of these established mechanisms is rather limited for ionic materials, in which intra-atomic electronic interactions are comparable to or even stronger than inter-atomic kinetic processes.
For example, these Fermi surface based mechanisms, in their conceptually reliable (weakly interacting) regime, require a sufficiently large density of states near the Fermi energy that many of these ionic materials lack.
Furthermore, the necessary dominant kinetic energy scale in these theories leads to a rather rigid ordering wavevector corresponding to the `nesting' of the Fermi surface that often does not match the experimentally observed ordering wavevector\,\cite{Shang2018Atomic-scale, Anjan2013, Xuetao2015, Inosov2008Fermi, Johannes2008Fermi, Aebi2001}.

The limitations of the standard theoretical framework are becoming increasingly apparent with recent significant improvements in experimental capability.
In NbSe$_2$, for instance, scanning tunneling microscopy reveals short-range charge correlations at temperatures far exceeding the transition temperature, $T_\mathrm{CO}$\,\cite{Arguello2014}. 
Such short-range correlations are further verified by X-ray diffraction\,\cite{Chatterjee2015} and inelastic X-ray scattering\,\cite{Moncton1975Study,Weber2011}.
Other signatures of short-range correlations, including phonon anomalies with broad peaks near the ordering wave vectors, have been observed in manganites\,\cite{Weber2010} and cuprates\,\cite{Reznik2006}.
Moreover, the differential pair distribution function analysis of resonant X-ray scattering reveals such local lattice distortions even \textit{prior to} the long-range coherence of charge density, magnetism, Kondo resonance, or superconductivity in many materials\,\cite{Petkov2020, Petkov2023Local, Petkov2023Charge, Andrea2024Local, Zafar2024Local, Bozin2023Crystallization}.
In TaS$_2$, Raman spectroscopy further confirms substantial lattice distortions\,\cite{Joshi2019}, while
ARPES studies reveal the persistence of the CDW energy gap above the transition temperature\,\cite{Zhao2017Orbital}. 
These observations all point to the existence of some microscopic mechanism at a relatively high energy scale that drives the short-range charge correlation in these ionic materials way beyond the temperature scale of the long-range ordering, below which these short-range correlations become coherent across the macroscopic domains.

With such $T_\mathrm{CO}$-exceeding strong short-range correlations, there is no reason to expect the ``normal'' phase above $T_\mathrm{CO}$ to be the weakly correlated Fermi liquid.
Indeed, many materials display ``non-Fermi liquid'' behaviors in spectroscopic and transport\,\cite{Borisenko2008Pseudogap,Mannella2005Nodal,Gurvitch1987Resistivity,Legros2019Planckian,
Mozaffari2024Universal} in the normal phase.
Such \textit{qualitative} departure from the Fermi liquid behavior immediately invalidates pictures involving instability of Fermi surfaces, including their strong coupling versions\,\cite{McMillan1977Microscopic, Varma1983Strong, Rossnagel2011On}.

An important clue for the identification of this microscopic mechanism is the observation that many materials exhibit multiple patterns of charge ordering under strain with similar volumes, such as $2 \times 2$, $4 \times 1$, and $3 \times 3$\,\cite{Shang2018Atomic-scale, Anjan2013}.
This suggests the key is some short-range physics with characteristic \textit{volume} that dictates the charge correlation, before the long-range order develops at lower energy scale\,\cite{Petkov2020, Petkov2023Local, Petkov2023Charge, Andrea2024Local, Zafar2024Local}.
Such a variety of patterns is inconsistent with the Fermi surface instability, which is rather insensitive to the strain owing to the Luttinger theorem\,\cite{Luttinger1960Ground-State, Luttinger1960Fermi}.
This indicates that such charge orders requires an alternative general \textit{volume}-sensitive paradigm applicable at higher energy- and shorter length-scale in these ionic materials.

In this work, we propose \textcolor{blue}{one} such general paradigm for charge order that is applicable to ionic systems hosting strong short-range correlations: the crystallization of long-lived Frenkel excitons\,\cite{Frenkel1931On}. 
We illustrate this mechanism in detail in the recently discovered kagome superconductors, CsV$_3$Sb$_5$\,\cite{Ortiz2019}, and show that it also offers a natural explanation for the distinct charge orders in ScV$_6$Sn$_6$.
Finally, we compare the generic characteristics of this new Frenkel exciton mechanism with those of the standard Fermi surface instability, which suggests the broad applicability of our theory as a general paradigm in understanding various physical behaviors associated with charge order in various ionic materials.

\section*{Insufficiency of current paradigms}

Currently, the set of qualitative distinct properties of unconventional charge orders has been well recognized to be beyond the standard paradigm of weakly unstable Fermi surface.
Consequently, one often phenomenologically regards these unusual properties as results of ``strong couplings''.
Such characterization is, however, insufficient for the essential electronic properties of a specific material, since it does not distinguish between the \textit{multiple} possible ``fixed points'' of emergent dynamics\,\cite{Anderson1959New, Anderson1972,Jiang2024Interation, Jiang2023Pressure} of strongly interacting systems, each with \textit{qualitatively distinct} 1) strong correlations and 2) slow emergent dynamics/fluctuations.
Analogous to ``non-Fermi liquids''\,\cite{Stewart2001NonFermiLiquid, Lee2018NonFermiLiquid}, it is not enough to state what the system is \textit{not} (a weakly unstable Fermi surface).
One needs to specify in what quantum state it really \textit{is} that produces the observed electronic properties besides a charge order.

\begin{figure}
\includegraphics[width=\linewidth]{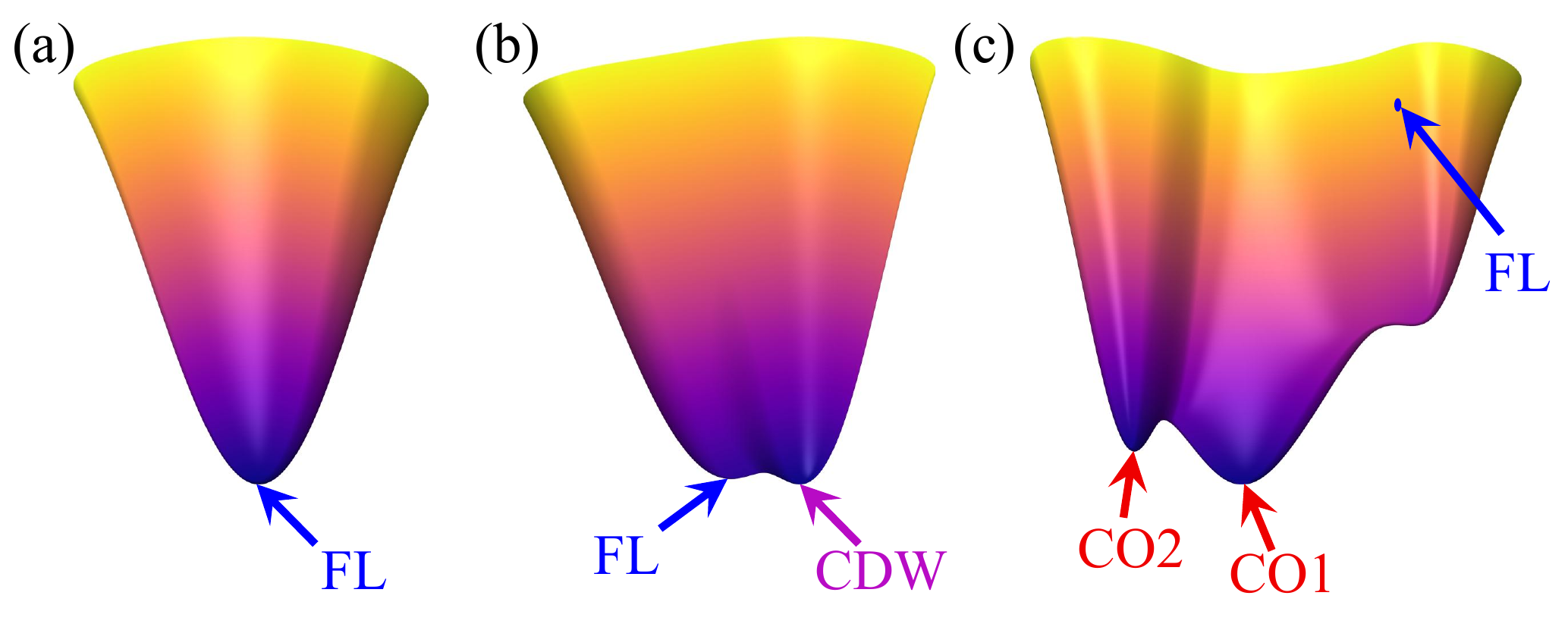}
\caption{\textit{Illustration of energy contours in an abstract parameter space}
(a) Fermi liquid (FL, blue arrow) state is well stabilized by kinetic energy.
(b) As a weak instability of the FL, the charge density wave (CDW, purple arrow) state is in the proximity of the FL state, along the most unstable direction (with the lowest gradient), typically detected via FL's susceptibility.
(c) In contrast, charge orders (COs, red arrows) result from strong short-range correlations and the emergent slow dynamic and are generically remote from the FL state.
(Notice the possibility of multiple stable CO states, exemplified by CO1 and CO2 states.)
The unstable direction of FL state thus does not reflect the structure or the ordering wavevectors of CO states.
}
\label{fig1}
\end{figure}

Indeed, very few of the above unconventional electronic properties were explicitly reproduced through the so-called ``strong coupling theories''\,\cite{McMillan1977Microscopic, Varma1983Strong}, other than a \textit{quantitatively} shorter coherent length\,\cite{McMillan1977Microscopic, Varma1983Strong} or a stronger instability\,\cite{Rossnagel2011On} with slight flexibility in the unstable wavevector.
This is to be expected, given that these theories still start with a Fermi surface under the influence of approximately treated stronger couplings, such that they are only reliable in the proximity of the starting Fermi liquid state [c.f. the CDW state in Fig.\,\ref{fig1}(b)].
In contrast, strongly correlated charge ordered states [such as the CO1 and CO2 states in Fig.\,\ref{fig1}(c)] with \textit{qualitatively distinct} physical properties are ``adiabatically disconnected'' from the Fermi liquid state and thus cannot be reliably accessed from the latter.
It is therefore timely to identify the most common correlated charge ordered states, whose dominant short-range correlation and emergent slow dynamics naturally allow the observed unusual properties, such as competing charge orders of drastically distinct wavevectors\,\cite{Zhao2021Cascade}, large spectral gap over transition temperature, $\Delta^*/T_{\text{CO}}\gg1$\,\cite{Ge2023Anharmonic}, first-order nature of the transition\,\cite{Subires2023Order-disorder}, absence of phonon softening\,\cite{Li2021Observation, Yubi2025Absence, Subires2023Order-disorder}, or non-Fermi liquid behavior\,\cite{Mozaffari2024Universal}, just to name a few.

\begin{figure*}
\centering
\includegraphics[width=0.95\textwidth]{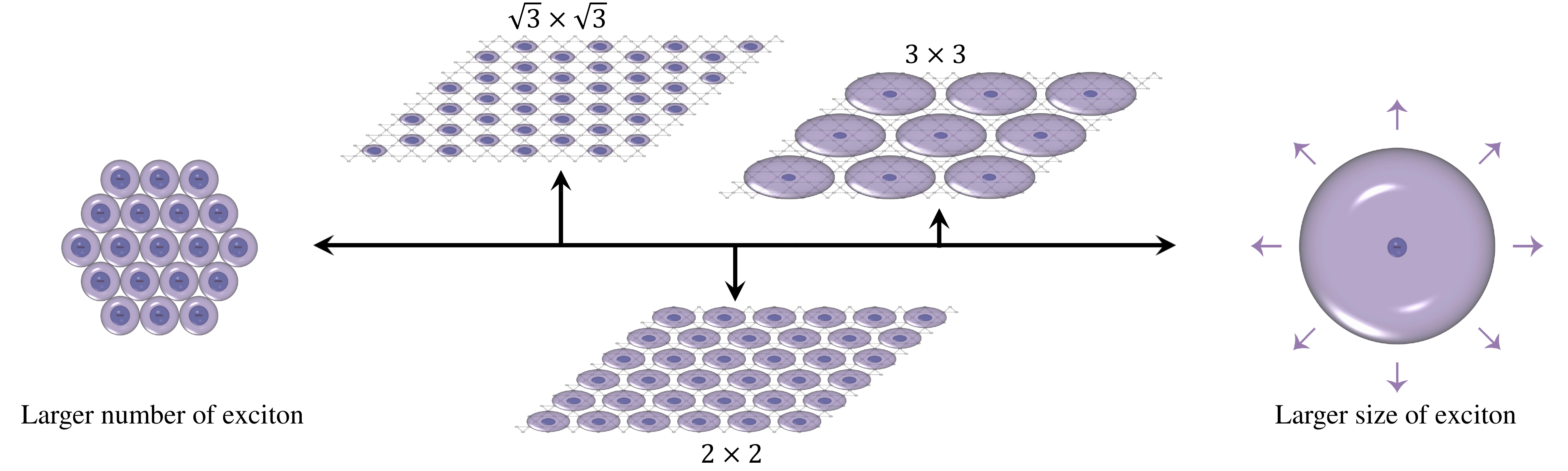}
\caption{\textit{Illustration of charge order via crystallization of Frenkel excitons.}
Given atomic orbital energies, two mechanisms compete to determine the lattice of exciton crystal.
Potential energy prefers higher density of small excitons (toward left), while excitons' internal kinetic energy tends to enlarge the exciton (toward right).
Constrained by the Pauli principle and emergent repulsive interaction of excitons' outer structure, a close-packed lattice results to optimize the competition.}
\label{fig2}
\end{figure*}

\section*{Charge order through crystallization of long-lived Frenkel excitons}

We build our proposed mechanism starting with the following scenario. First, we consider ionic systems that are energetically unstable when the effective valence of ions is uniformly integer across the sample.
That is, we consider systems that can lower their energy by allowing some degree of local charge transfer between neighboring ions, effectively generating tightly bound Frenkel excitons to stabilize the system.
Here, the exact microscopic mechanism, such as electronic or electron-lattice interactions, is unessential as long as its energy scale is large.
Second, we focus on cases with charge fluctuations that are not strong enough to dynamically reverse the charge transfer, namely systems with long-lived local Frenkel excitons having low recombination rate.

Under these conditions, the \textit{strong} short-range charge correlation within the tightly bound Frenkel excitons is known\,\cite{Anderson1959New,Lee2013First-Principles} to generically suppress spatial overlap of the low-energy emergent objects (\textit{i.e.} excitons).
Consequently, the system would spontaneously generate some number of Frenkel excitons according to the optimal balance between the potential and kinetic energies of the system.
As illustrated in Fig.\,\ref{fig2}, in the limit where the potential energy of exciton formation dominates (toward left), the system prefers a large number of small excitons.
In the opposite regime where the kinetic energy of the shorter-time \textit{internal} dynamics of excitons is strong (toward right), the system prefers fewer excitons of larger size.
In parallel, the internal electron-hole structure of the excitons implies that their outer regions are subject to the Pauli exclusion principle and Coulomb interaction, leading to an effective ``hard core'' for each exciton. Combined, all these considerations lead to a final exciton configuation that provides an optimal balance between a higher density of smaller excitons and a smaller number of larger excitons.
Naturally, at low enough temperature such optimization can be reached through the formation of a close-packed periodic lattice of excitons.
Such crystallization of long-lived Frenkel excitons therefore displays the charge order observed in the experiments.

Frenkel exciton crystallization is a fundamentally different mechanism for charge order compared to Fermi surface instabilities, and therefore one expects qualitatively distinct slow dynamics and corresponding phenomenology.
Below, we first illustrate this generic scenario using the kagome superconductor CsV$_3$Sb$_5$ as an example, and then discuss the distinguishing features of charge order driven by Frenkel exciton crystallization in contrast to those of the standard mechanism of Fermi surface instability.

\section*{Demonstration using kagome superconductor CsV$_3$Sb$_5$}

Charge order has recently been discovered in kagome metals $\rm AV_3Sb_5$ $\rm (A=K, Rb, Cs)$\,\cite{Ortiz2019,Jiang2021}, whose structure consists of an unusual combination of honeycomb, triangular, and kagome sublattices hosting Dirac cones in the band dispersion.
These materials became the subject of intense research interest with the discovery of superconductivity\,\cite{Ortiz2020CsV3Sb5,Ortiz2021superconductivity,Yin2021Superconductivity}, and multiple superconducting phases were later found under pressure\,\cite{Du2021Pressure, Zhu2022Double-dome, Lin2022Multidome}.
Furthermore, these materials display anomalous Hall and Nernst effects\,\cite{Yang2020Giant, Chen2022Anomalous, Mi2022Multiband, Yu2021Concurrence, Zhou2022Anomalous} and anomalous transport and magnetic behaviors\,\cite{Nguyen2022Electronic} which cast doubt on the applicability of Fermi liquid theory and consequently on the Fermi surface instability mechanism for the description of their emergent phenomena.

Focusing on their unconventional charge order, long-range order occurs ranging from $80$ to $100$\,K in the $\rm AV_3Sb_5$ family\,\cite{Jiang2021, Li2021Observation, Uykur2022Optical, Liang2021Three-Dimensional, Zhou2021Origin, Ortiz2021Fermi}.
Most experiments find the charge ordering phase transition to be likely first-order\,\cite{Subires2023Order-disorder, Park2023Condensation, Ratcliff2021Coherent}.
Interestingly, the charge order is found to be accompanied by a simultaneously broken rotational symmetry\,\cite{Li2022Rotation,Zhao2021Cascade,Miao2021Geometry, Wenzel2022Optical, Xu2022Three, Asaba2024Evidence, Guo2024Correlated, Wu2022Simultaneous} and sometimes time-reversal symmetry\,\cite{Khasanov2022Time-reversal, Hu2022Time-reversal, Jiang2021Unconventional, Shumiya2021Intrinsic, Wang2021Charge, Feng2021Chiral, Wu2021Nature, Yu2021Evidence, Guo2022Switchable, Wang2021Electronic}.
Most notably, alongside the prominent $2\times 2$ lattice reconstruction, a clear signature of $4\times 1$ order is also observed\,\cite{Li2022Discovery,Shumiya2021Intrinsic, Zhao2021Cascade, Wang2021Charge}, which is inconsistent with the nesting wavevector of the Fermi surface of the system.
This family therefore serves as a good candidate to explore our proposed scenario of Frenkel exciton crystallization.

\begin{figure*}
\centering
\includegraphics[width=0.82\textwidth]{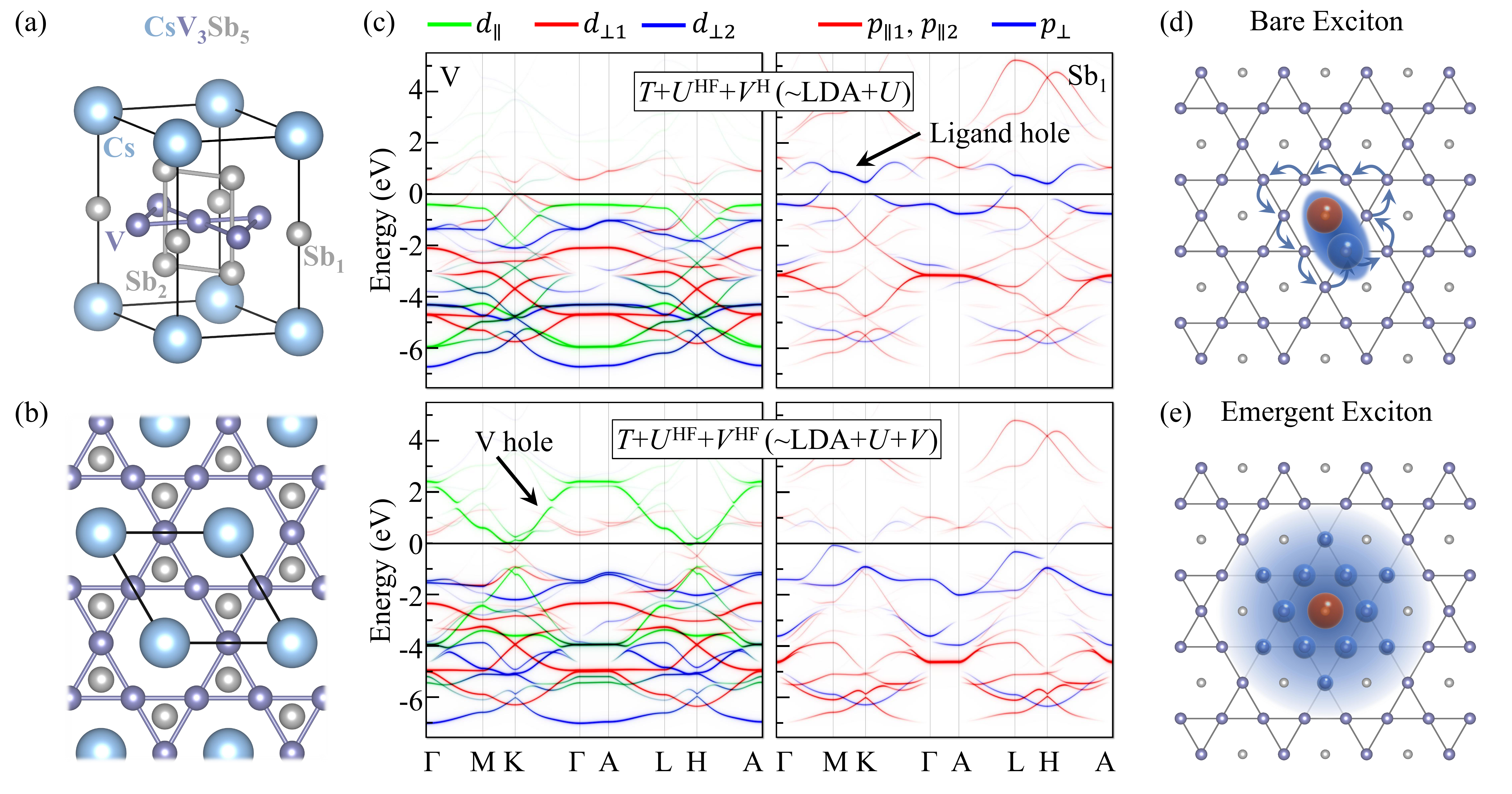}
\vspace{-0.3cm}
\caption{\textit{Formation of long-lived Frenkel excitons.}
(a) and (b) illustrate the lattice structure of CsV$_3$Sb$_5$.
(c) The LDA+$U$ band structure of the majority spin in the upper panel gives a uniform charge distribution, corresponding to an effective $\ket{d^3d^3d^3\underline{L}}$ configuration.
Improving it with additional inter-atomic interaction $V$ transfer the ligand hole from the blue Sb$_1$-$p_\perp$ orbital (upper right) to the green in-plane V-$d_\parallel$ orbital (lower left).
The resulting $\ket{d^2d^3d^3}$ configuration corresponds to the formation of a \textit{bare} Frenkel exciton as a tightly bound particle(red)-hole(blue) pair whose recombination is forbidden by their distinct $z$-parity, as illustrated by an \textit{instantaneous} snap shot in (d).
(e) Upon absorbing the short-time dynamics of the holes, the \textit{emergent} long-lived Frenkel excitons would contain a electron in the center and a hole-cloud around it in the same plane.
}
\label{fig3}
\vspace{-0.3cm}
\end{figure*}

\subsubsection*{Existence of long-lived excitons}

Using CsV$_3$Sb$_5$ as a representative example\,\cite{supplementary}, we start by examining whether the conditions for long-lived Frenkel excitons required for our proposed mechanism are satisfied.
In a previous LDA+$U$ study\,\cite{Jiang_localmoment} with realistic intra-atomic interaction strength $U=0.6$\,Ry and $J=0.06$\,Ry, it was found that the out-of-plane ligands Sb$_2$ [c.f. Fig.\,\ref{fig3}(a)(b) and Supplementary\,\cite{supplementary}] develop covalent bonds across the planes and contribute to \textit{massless} Dirac-type carriers.
On the other hand, the in-plane ligands (surrounded by six V ions) are strongly ionic, each with a hole in their $p_\perp$ orbitals.
Indeed, as shown in the upper panels of Fig.\,\ref{fig3}(c), the presence of a ligand hole, $\underline{L}$, is made clear by the half-filled blue Sb$_1$ $p_\perp$ band in both spin-majority and minority (c.f. SI\,\cite{supplementary}) channels of the in-plane Sb ions in the right panel.
The spin-majority channel of V shown in Fig.\,\ref{fig3}(c) also indicates that \textit{locally} V ions are in an \textit{effective} V$^{2+}$ $d^3$ spin-$3/2$ configuration with fully spin-polarized effective $t_{2g}$ orbitals, as expected from the Hund's rules.
Given that correlations, such as those driving exciton formation, are most pronounced in massive orbitals, the relevant low-energy degrees of freedom in these materials are therefore the V $t_{2g}$ orbitals and Sb$_1$ $p_\perp$ orbitals, in the $\ket{d^3d^3d^3\underline{L}}$ configuration.

To properly describe the binding of Frenkel excitons, we extend the calculation to LDA+$U$+$V$ that includes also the nearest neighbor (self-interaction-free) Coulomb repulsion $V$:
\begin{align}
        H^{\text{eff}} &= \sum_{i,m,\nu}\epsilon_mc^\dagger_{im\nu}c_{im\nu}+\sum_{i,i^\prime,m,m^\prime,\nu}t_{im,i^\prime m ^\prime}c^\dagger_{im\nu}c_{i^\prime m^\prime \nu}    \label{eq1}
        \\ &+\frac{1}{2}\sum_{i,m,m^\prime,m^{\prime\prime},m^{\prime\prime\prime},\nu,\nu^\prime} U_{m m^{\prime\prime} m^\prime m^{\prime\prime\prime}} c^\dagger_{im\nu}c^\dagger
        _{im^{\prime\prime}\nu^\prime}c_{im^{\prime\prime\prime}\nu^\prime}c_{im^\prime\nu} \nonumber \\ 
        &+\frac{1}{2}\sum_{i, i^\prime,m,m^\prime,\nu,\nu^\prime} V_{im, i^\prime m^\prime} c^\dagger_{im\nu}c^\dagger_{i^\prime m^{\prime}\nu^\prime}c_{i^\prime m^{\prime}\nu^\prime}c_{im\nu}. \nonumber
\end{align}
in addition to the one-body orbital energy $\epsilon$, kinetic hopping strength $t$, and intra-atomic interaction $U$ among the V-$d$ orbitals, denoted by creation $c^\dagger_{im\nu}$ and annihilation $c_{im\nu}$ operators of orbitals $m$ and spin $\nu$ within unit cell $i$\,\cite{supplementary}.

The lower panels of Fig.\,\ref{fig3}(c) display the resulting orbital-projected one-body spectral functions obtained from a self-consistent Hartree-Fock calculation under a realistic inter-atomic interaction $V=0.07$\,Ry.
The results, which are unfolded\,\cite{Wei2010Unfolding} to the original unit cell from a $2\times 2$ supercell that allows the occurrence of zero to four Frenkel excitons, reveal that the original $\ket{d^3d^3d^3\underline{L}}$ configuration is not energetically the most stable, consistently with the first condition of our scenario.
Instead, relocating a ligand hole from the blue Sb$_1$-$p_\perp$ orbital (upper right panel) to the green V-$d_\parallel$ orbital (lower left panel), effectively a $\ket{d^2d^3d^3}$ ionic configuration, is energetically favorable.
From a chemical point of view, this is not surprising given that Sb with a full $p$-shell should have some energetic advantage.
Relative to the uniform $\ket{d^3d^3d^3\underline{L}}$ reference state, this state corresponds to the spontaneous occurrence of tightly bound Frenkel excitons, with the (electron, hole)-pair residing in neighboring (Sb$_1$-$p_\perp$, V-$d_\parallel$) orbitals shown in Fig.\,\ref{fig3}(d).

Importantly, annihilation of this \textit{bare} Frenkel exciton via charge fluctuations that recombine the electron and hole is \textit{forbidden} by symmetry.
Specifically, the Sb$_1$-$p_\perp$ orbital that hosts the electron has odd $z$-parity and thus has \textit{no} charge motion to the even $z$-parity V-$d_\parallel$ orbital that hosts the hole.
(For example, our DFT computed $t_{im,i^\prime m^\prime}=0$ between these two orbitals.)
As a consequence, this particular exciton exhibits exceptional longevity, satisfying the second condition of our scenario.
Once exist, these bare Frenkel excitons cannot undergo dynamical number fluctuations, a crucial characteristic for stabilizing long-lived phenomena such as charge order.

For such long-lived phenomena, it is physically more intuitive and convenient to absorb the rapid dynamics into the internal structure of larger \textit{emergent} Frenkel excitons associated with longer timescales.
In this example, the electrons in the Sb$_1$-$p_\perp$ orbital have restricted out-of-plane kinetics, given the highly ionic Cs ions right above and below them.
Similarly, the dominant hopping for holes in V-$d_\parallel$ are also in-plane due to its spatial distribution. 
Furthermore, the in-plane kinetic processes of the V-$d_\parallel$ orbitals ($t_{dd}\sim 600$\,meV, \textit{c.f.} Eq.\,\ref{eq2} below) are more than an order of magnitude more efficient than those of the Sb$_1$-$p_\perp$ orbitals ($t_{pp}\sim 20$\,meV).
Consequently, as illustrated in Fig.\,\ref{fig3}(d), a hole in the former would rapidly orbit around a localized electron in the latter.
As shown below in the case study, at longer timescales the emergent long-lived Frenkel exciton [obtained via local diagonalization, $\tilde{b}^\dagger\equiv \mathcal{U}^\dagger b^\dagger \mathcal{U}=\mathcal{U}^\dagger (p^\dagger d) \mathcal{U}$, via a unitary $\mathcal{U}$, see below] would contain an electron at the Sb$_1$ site and a hole cloud around it in the same plane, as illustrated in Fig.\,\ref{fig3}(e), with a heavily suppressed exciton kinetic strength, $t_{ex}\ll t_{pp}$, easily defeatable by the inter-exciton repulsion.
Such emergent excitons align perfectly with the experimental real-space pattern of charge profile and the local density of states\,\cite{Jiang2021Unconventional}.

\subsubsection*{Crystallization for optimal energy}

Given that both conditions of our proposed scenario are fulfilled in CsV$_3$Sb$_5$, we proceed to determine the optimal balance between the kinetic energy that favors larger excitons and the potential energy that favors a larger number of excitons.
First recall that it is the high-energy exclusive (or strongly repulsive) charge \textit{correlation} between the outer structure of the excitons [similar to the double-occupation constraint in the $t$-$J$ model\,\cite{Anderson1959New}] that forces these two mechanisms to compete.
(Without this key correlation, as soon as charge transfer is energetically favored, the exciton population would unphysically grow too large, just as in Hartree-Fock mean-field treatments.)
Under such exclusive correlation,  to achieve the optimal energy balance (c.f. Fig.\,\ref{fig2}) the nearly immobile emergent excitons must maximally fill the space and in turn block the propagation of each other as in Mott insulators\,\cite{Imada1998} or Wigner crystals\,\cite{Wigner1934}.

Therefore, the low-energy many-body states (indexed by $I$) with $N_b$ excitons must be predominantly,
\begin{equation}
  \ket{\Psi_I; N_b}\sim\prod^{N_b}_j\tilde{b}^\dagger_j\ket{\mathrm{FL}}\sim\prod^{N_b}_j\tilde{b}^\dagger_j\prod_{k<k_\mathrm{F}} c^\dagger_k\ket{0},
  \label{CO}
\end{equation}
with $N_b$ optimized emergent excitons, $\tilde{b}^\dagger_j$, located at  (ordered or disordered) sites $\{j\}_I$, each residing within a region $\Omega_j$ spatially constrained by its surrounding neighbors (c.f. Fig.4).
Correspondingly, the internal dynamics of the emergent exciton (EE) with an electron $p^\dagger_j$ in the Sb$_1$-$p_\perp$ orbital at site $j$ and a \textit{hole}, $d^\dagger_{im}$, in the V-$d_\parallel$ orbitals at site $(i,m)$ is described by,
\begin{align}
    H^\mathrm{EE} &= \sum_{\{i,m; i^\prime,m^\prime\}\in \Omega(j)}t_{im,i^\prime m^\prime}d^\dagger_{im}d_{i^\prime m^\prime}     \label{eq2}
    \\ &+ \epsilon\sum_{\{i,m\}\in \Omega(j)} d^\dagger_{im}d_{im} - V^\mathrm{B}\sum_{\{i,m\}\in \mathrm{NN}(j)}  d^\dagger_{i m}p^\dagger_{j}p_{j}d_{im},\nonumber
\end{align}
before getting dressed by other quantum fluctuations.
(Here the spin indices are omitted for simplicity.)

In principle, parameters of such an highly emergent effective description are expected to be renormalized from Eq.~\ref{eq1}.
Nonetheless, for a simple illustration of the trend in Fig.~\ref{fig2}, we proceed with the LDA+$U$ extracted parameters: the orbital energy difference between Sb$_1$-$p_\perp$ and V-$d_\parallel$ orbitals, $\epsilon$, and internal kinetic hopping of the hole $t_{im,i^\prime m^\prime}$.
The physical trends are then demonstrated through the ground-state energies per formula unit within a range of \textit{effective} particle-hole binding energy, $V^\mathrm{B}\in [0,2]$ eV, that alters the essential formation energy of the nearest neighboring ``bare'' excitons, $E^\text{F} = \epsilon-V^\mathrm{B}$.

\begin{figure*}
\centering
\includegraphics[width=\textwidth]{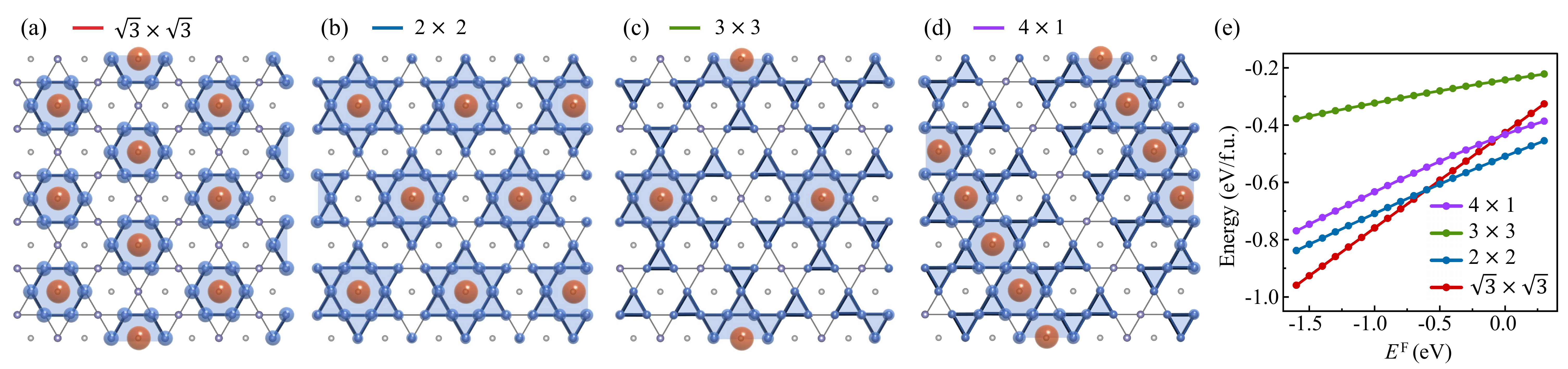}
\vspace{-0.3cm}
\caption{\textit{Energetic optimization of exciton crystallization.} 
Consider four close-packed lattices of rotational symmetric emergent excitons with different size (a) $\sqrt{3} \times  \sqrt{3}$, (b) $2 \times  2$, (c) $3 \times 3$, and (d) $4 \times  1$  under various strength of binding energy with fixed orbital energy and kinetic strengths.
(e) The total energy per formula unit (f.u.) indicates that the $\sqrt{3} \times  \sqrt{3}$ lattice is energetically favored at strong binding (lower formation energy $E^\mathrm{F}$), while with weaker binding ($E^\mathrm{F} > -0.7$ eV) the $2 \times  2$ lattice takes over to benefit from the kinetic energy at lower density.
Also notice the systematically similar energies for lattices of similar volume/area, $2\times 2$ and $4\times 1$, due to their comparable number of internal kinetic processes, a characteristic distinct from the nesting consideration of Fermi surface instability.}
\label{fig4}
\vspace{-0.3cm}
\end{figure*}

The considered cases include $\sqrt{3} \times \sqrt{3}$, $2 \times  2$, and $3 \times 3$ shown in Figs.\,\ref{fig4}(a)-(c), corresponding to exciton densities of $3^{-1}$, $4^{-1}$, and $9^{-1}$ per formula unit, respectively.
Figure\,\ref{fig4}(e) then gives the resulting energy comparison for three close-packed lattices of rotation-symmetric emergent excitons with different sizes.
While higher exciton densities lead to greater potential energy gains, this advantage is offset by reduced kinetic energy benefits.

This trade-off becomes particularly evident in the $3 \times 3$ lattice, where the low density of large excitons results in both inefficient kinetic energy gain and insufficient binding-associated potential energy, rendering this configuration energetically unfavorable. 
In contrast, the $2 \times  2$ and $\sqrt{3} \times \sqrt{3}$ lattices, with their higher densities of smaller excitons, achieve a more optimal balance between internal kinetic energy and overall potential energy.
Specifically, following the trend of the general picture in Fig.\,\ref{fig2}, under a strong effective binding, $V^\mathrm{B}$ (lower $E^\mathrm{F}$), the higher density of smaller excitons in the $\sqrt{3} \times \sqrt{3}$ lattice is energetically optimal.
For weaker binding, $E^\mathrm{F} > -0.7$ eV, it is more advantageous to have slightly lower density of larger excitons to benefit from the kinetic energy and form a $2\times 2$ exciton lattice, in agreement with the experimentally observed $2 \times  2$ charge order\,\cite{Jiang2021Unconventional}.

\subsubsection*{Out-of-plane stacking}

\begin{figure*}
\centering
\includegraphics[width=0.85\textwidth]{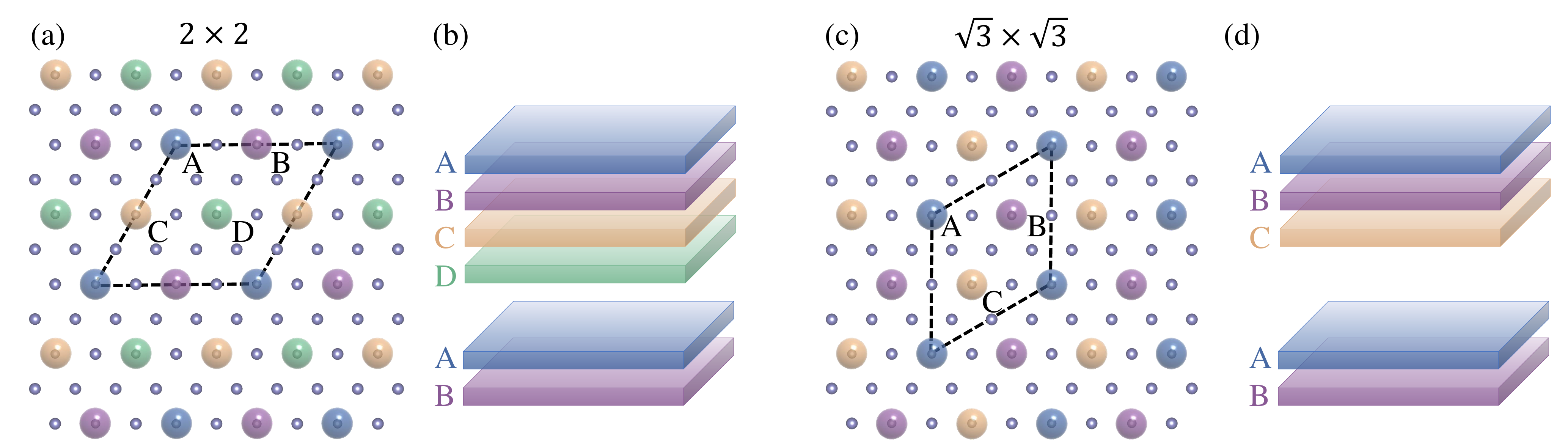}
\caption{ $z$-axis stacking in the kagome superconductors. (a)(b) show the ABCD 2x2x4, or imperfectly 2x2x2 in CsV$_3$Sb$_5$ and (c)(d) show the ABC $\sqrt{3}\times \sqrt{3}\times 3$, or imperdtly AB $\sqrt{3}\times \sqrt{3}\times 2$ in ScV$_6$Sn$_6$. } 
\label{fig5}
\vspace{-0.3cm}
\end{figure*}

Once the optimal 2-dimensional lattice of the emergent excitons is determined via optimizing the in-plane physics, the lower-energy out-of-plane stacking of weakly coupled 2-dimensional layers naturally follows the classical stacking of planes of charged hard spheres.
As shown in Fig.\,\ref{fig5}(a)(b), for a $2\times 2$ lattice of emergent excitons, the optimal stacking (with lowest Coulomb energy) would be a 4-period ABCD stacking with excitons centered around one of the four available Sb$_1$ sites in each layer, corresponding to a $2\times 2\times 4$ charge order, in agreement with the experimental observations\,\cite{Ortiz2021Fermi,Wu2022Charge, Luo2022Possible}.
Alternatively, the imperfect 2-period ABAB ones (with only slightly higher interlayer Coulomb energy), may also occur as observed experimentally\,\cite{Li2021Observation,Subires2023Order-disorder, Liang2021Three-Dimensional, Song2022Orbital}.

\subsubsection*{First-order finite-temperature phase transition}

The focus of this manuscript is on the crystallization of Frenkel excitons as a generic class of quantum states of matter with charge order.
Nonetheless, given their large energy scale, naturally the microscopic dynamical origins of the Fenkel excitons must be insensitive to thermal fluctuations of the room temperature scale ($\sim 30$ meV) and the corresponding short-range physics must persist \textit{far beyond} the observed transition temperature, $T_\mathrm{CO}$.
Therefore, the phase transition must be associated with a loss of long-range ``phase coherence'' of the exciton crystal, instead of the number fluctuation of the Frenkel excitons.

Intuitively, a general entropy consideration would dictate melting of the exciton crystal  at some temperature, at which the long-range charge order vanishes.
A careful study of such a finite-temperature phase transition is a serious study on its own involving determining the nature of the transition and,  for the continuous ones, the corresponding ``universality class''\,\cite{WilsonKogut1974}.
Here, for the prototypical example of CsV$_3$Sb$_5$, we only briefly demonstrate its first-order nature with a reasonable temperature scale.

First, note that due to the rather high exciton formation energy, in Fig.\,\ref{fig4} the energy differences between crystal states of different exciton density are of the order of 200 meV, much larger than the 30 meV scale of the transition temperature.
Therefore, the relevant thermal fluctuations for the crystal melting must be dominated by states with nearly identical exciton density but with dislocated domains of locally ordered structures.
For quasi-2D systems, the energy of such states is thus dominated by the length of the domain walls\,\cite{Peierls1936, Griffiths1964}.
Employing the same computation as the ordered state via Eq.\,\ref{eq2}, we find the energy density per unit length of the dominant domain walls to be $\sim 60$ meV\,\cite{supplementary}.
Via the Peierls contour consideration of domain walls\,\cite{Peierls1936, Griffiths1964}, the estimated melting temperature, $k_\mathrm{B} T_\mathrm{CO} \sim 37$ meV\,\cite{supplementary}, is qualitatively consistent with the observation.
(Physically renormalized parameters in Eq.\,\ref{eq2} should quantitatively lower $T_\mathrm{CO}$ even closer to the observation.)

To further confirm in this system the well-known first-order nature of a melting transition of lattices\,\cite{Alexander1978}, we focus on states with disordered lattice under a fixed exciton density of the ordered state.
Then, upon mapping the energy differences between states of different structures to effective short-range inter-exciton repulsions, we simulate the melting of Frenkel excitons  near $T_\mathrm{CO}$ via a Monte Carlo algorithm\,\cite{supplementary}.
Within the capability of a finite-size simulation, the resulting internal energy already displays an abrupt change near $T_\mathrm{CO}\sim 38$ meV, accompanied by a sharp peak in the specific heat.
Together with the double stable structures in the Monte Carlo histogram, these results reproduce the observed first-order nature of the transition, at the temperature similar to the above domain wall-based estimation.

\section*{Generic features of charge order driven by Frenkel exciton crystallization}

Having illustrated the newly proposed mechanism of Frenkel exciton crystallization as the driving mechanism for charge order in CsV$_3$Sb$_5$, we next explore the \textit{universal } qualitative characteristics of charge order resulting from this \textit{strongly correlated} mechanism. 
Strikingly, these features, which differ fundamentally from those associated with conventional Fermi surface instabilities, show excellent agreement with various anomalous experimental findings in correlated materials exhibiting charge order.

The scenario of Frenkel exciton crystallization starts with high-energy short-range local charge correlations, which develop into coherent long-range order at lower energy scales.
The dominant physics in this scenario, namely strong local electron/lattice correlations, is fundamentally distinct from that of the standard Fermi surface instability mechanism\,\cite{Tan2021Charge, Li2021Observation, Jiang2021Unconventional, Denner2021Analysis, Lin2021Complex, Kang2022Twofold}, characterized by coherent kinetics perturbed by weak interactions or electron-phonon coupling\,\cite{Gutierrez2024Phonon, Alkorta2025symmetry}.
Therefore, one expects its characteristics
to be qualitatively distinct from the standard lore.
Below, we systematically examine these unique features across energy scales, with key representative properties summarized in Tab.\,\ref{tab1}.

First, the Frenkel exciton crystallization scenario accepts remarkable flexibility in both spatial patterns and ordering wavevectors.
The internal kinetic energy gain of emergent excitons depends primarily on the number of internal kinetic processes, making it more sensitive to the size than the shape of excitons.
This size-dependence implies similar energies for states with comparable exciton densities, regardless of their different lattice configurations.
As a demonstration, consider a less symmetric $4\times 1$ excitonic lattice shown in Fig.\,\ref{fig4}(d) with identical volume to the $2\times 2$ lattice in (b).
Indeed, Fig.\,\ref{fig4}(e) shows that the former is only slightly higher in energy, while sharing the same overall trend.
Therefore, unlike the rather robust nesting wavevectors associated with instabilities of the Fermi surface, emergent excitons may crystallize into multiple lattice structures with distinct ordering wavevectors but of similar exciton density.
It is therefore more likely for the Frenkel exciton crystallization scenario to develop charge order with lower rotational symmetry in both the local lattice deformation and the ordering wavevectors\,\cite{Kivelson2003How}, as frequently observed in kagome metals (KV$_3$Sb$_5$\,\cite{Li2022}, RbV$_3$Sb$_5$\,\cite{Shumiya2021Intrinsic}, CsV$_3$Sb$_5$\,\cite{Li2022Discovery,Wang2021Charge}), and cuprates (Bi$_{2-y}$Pb$_y$Sr$_{2-z}$La$_z$CuO$_{6+x}$\,\cite{Wise2008Charge-density-wave}, Ca$_{2-x}$Na$_x$CuO$_2$Cl$_2$\,\cite{Hanaguri2004A}, La$_{2-x}$Sr$_x$CuO$_4$\,\cite{Ando2002Electrical}, YBa$_2$Cu$_3$O$_y$\,\cite{Ando2002Electrical}).
Similarly, a change in the ordered excitonic lattice structure can be easily triggered by applied pressure, strain, or fields, as observed in transition metal dichalchogenides (NbSe$_2$\,\cite{Shang2018Atomic-scale, Anjan2013}, TiSe$_2$\,\cite{Novello2017Stripe}, TaS$_2$\,\cite{Tsen2015Structure}, VSe$_2$\,\cite{Duvjir2021Multiple}), or even under lower temperature, as in nickelates (Ba$_{1-x}$SrNi$_2$As$_2$\,\cite{Lee2021Multiple}), kagome superconductors (LaRu$_3$Si$_2$\,\cite{Ma2025Correlation}), transition metal dichalchogenides (NbTe$_2$\,\cite{Bai2023Realization}, TaS$_2$\,\cite{Tsen2015Structure}), and manganates (Bi$_{1-x}$Sr$_{x-y}$Ca$_y$MnO$_3$\,\cite{Schnitzer2025Atomic-Scale}).

Second, given the rapid decrease of the \textit{additional} internal kinetic energy gain upon further increasing the exciton size, this scenario typically disfavors very large emergent excitons.
That is, the corresponding charge order typically avoids long-period modulations (small wavevectors.)
Similarly, in the quantum (low-temperature) limit, charge-order commensurate with the atomic lattice is typically energetically favored.
Consequently, when accumulation of stress reaches a critical value, a discommensuration\,\cite{McMillan1976Theory} is a more natural choice adjust the overall periodicity while still respecting the commensurate high-energy local structure.
Correspondingly, over the length scale of the ordered period, the spatial density profile is defined mostly by atomic orbitals, rather than by harmonic waves as in charge density waves.
Furthermore, owing to its internal charge redistribution, an immobile Frenkel exciton is always dressed by nearby ions' displacement as a charge-neutral polaron\,\cite{Bozin2023Crystallization}.
Such locally commensurate ``polaronic'' structure has, in fact, been overwhelmingly observed in correlated materials, such as the transition metal dichalchogenides (TaS$_2$\,\cite{Tsen2015Structure}, TaSe$_2$\,\cite{Giambattista1990Scanning}, VSe$_2$\,\cite{Giambattista1990Scanning}, NbSe$_2$\,\cite{Moncton1977Neutron}), manganates (Bi$_{1-x}$Sr$_{x-y}$Ca$_y$MnO$_3$\,\cite{Schnitzer2025Atomic-Scale}, La$_{1-x}$Ca$_x$MnO$_3$\,\cite{Chen1996Commensurate}, Pr$_{1-x}$Ca$_x$MnO$_3$\,\cite{Chen1999Anomalous}), nickelates (La$_{1-x}$Sr$_x$NiO$_2$\,\cite{Rossi2022A}), cuprates (YBa$_2$Cu$_3$O$_y$\,\cite{Vinograd2021Locally}), kagome metals (KV$_3$Sb$_5$\,\cite{Jiang2021}, RbV$_3$Sb$_5$\,\cite{Li2021}, CsV$_3$Sb$_5$\,\cite{Song2021,Li2021}, ScV$_6$Sn$_6$\,\cite{Lee2024Nature} ), iridium alloys (IrTe$_2$\,\cite{Kim2015Origin}, Lu$_5$Ir$_4$Si$_{10}$\,\cite{Becker1999Strongly}, Lu$_2$Ir$_3$Si$_5$\,\cite{Sangeetha2015Multiple}), and cobaltates (CoO\,\cite{Negi2021Coexisting}, Na$_x$CoO$_2$\,\cite{Foo2014}).

\begin{table*}
\centering
\caption{
Comparison between Fermi surface instability and Frenkel exciton crystallization mechanisms for charge order.
}
\begin{tabular}{c||c|c} \hline
& \textit{Fermi surface instability} & \textit{Frenkel exciton crystallization} \\  \hline \hline
\textit{Building block} & Long-range coherence & Short-range correlation \\ \hline
\textit{Key origin} & Density of state at $E_\mathrm{F}$ & Short-range coulomb energy \\ \hline
\textit{Dominant physics} in&Momentum & Position \\ \hline
\textit{Constraint in slow dynamics} & None & Short-range exclusion \\ \hline
\textit{Sensitive to} & Periodicity (wavevector) & Supercell size (density) \\ \hline
\textit{Typical period} & $\pi/k_\mathrm{F}$ (Inconmensurate) & Commensurate / Discommensurate \\ \hline
\textit{Role of FS nesting} & Driving FS instability & Reducing long-range phase fluctuation \\ \hline
\textit{Density modulation} &Harmonic & Atomic \\ \hline
\textit{Lattice distortion} & Harmonic & Polaron-like  \\ \hline
\textit{Spin texture} &No & Possible  \\ \hline
\textit{Typical local symmetry} & High & High / Low\\ \hline
\textit{Leading fluctuation} &Amplitude (pairing) & Phase (coherence) \\ \hline
\textit{Energy scale of $T_{\text{CO}}$} & Interaction $\sim\Delta$ & Long-wavelength phase stiffness $\ll\Delta^*$\\ \hline
\textit{At $T \approx  T_{\text{CO}}$} & Vanishing local order & Exciton crystal melting\\ \hline
\textit{Phase transition} & Second-order & First- / Second-order\\ \hline
\textit{Phonon frequency near $T_{\text{CO}}$} & Softens to zero & Possibly remains finite\\ \hline
\textit{At $T>T_{\text{CO}}$} & Nearly uniform density & Strong local deformation\\ \hline
\textit{Electronic structure} & Fermi liquid (FL) & non-FL / Strongly correlated FL\\ \hline
\end{tabular}
\label{tab1}
\end{table*}

Third, since the dominant physics of the Frenkel exciton crystallization scenario are the interaction-driven local correlations, for charge-ordered states with low itinerant carrier densities the Fermi surface nesting would play a weaker role in driving the charge order.
Such \textit{disrespect} to nesting wavevectors has in fact been observed in many ionic materials\,\cite{Xuetao2015}, for example the transition metal dichalchogenides (NbSe$_2$\,\cite{Shang2018Atomic-scale, Anjan2013, Xuetao2015, Inosov2008Fermi}, TaSe$_2$\,\cite{Inosov2008Fermi}, TiSe$_2$\,\cite{Aebi2001}, Cu$_x$NbS$_2$\,\cite{Inosov2008Fermi}).

On the other hand, it is known that with higher density of itinerant carriers in ordered states, long-range quantum fluctuations can disrupt the long-range phase coherence of the order\,\cite{Ruderman1954Inderect,Yuting2015Itinerancy-Enhanced}, resulting in a quantum unordered phase\,\cite{Tan2022Stronger,Yuting2015Itinerancy-Enhanced}.
In that case, a rekindled long-range order\,\cite{Hou2023Chemical} may re-emerge in the presence of chemical or structural disorder, or another ordering wavevector of similar exciton density can take over.
From this consideration, a reasonably nested normal-state Fermi surface compatible with the period of the exciton crystal, \textit{rather than} driving the order, can still be very helpful in suppressing long-range quantum fluctuations of the ordering.
This coherence perspective of nesting is perhaps more relevant than the traditional nesting picture in ionic materials, such as transition metal dichalchogenides (TaTe$_2$\,\cite{Lin2022Evidence}, NbSe$_2$\,\cite{Inosov2008Fermi}), kagome metals (CsV$_3$Sb$_5$\,\cite{Kang2022Twofold}), nickel pnictide superconductors [Ba(Ni$_{1-x}$Co$_x$)$_2$As$_2$\,\cite{Lee2019Unconventional}], and cuprates (Bi$_2$Sr$_{2-x}$La$_x$CuO$_{6+\delta}$\,\cite{Comin2014Charge}).

Fourth, since the formation of local emergent excitons is driven by higher-energy local interactions, these excitons can sometimes carry a spin-1 texture, namely opposite spins for the electron and hole components.
In that case, the crystallization of such emergent excitons would \textit{simultaneously} order in the spin channel (a \textit{single} order parameter having both charge and spin texture.)
For example, along the direction of an anti-ferromagnetic order for excitons, the magnetic order would have a period associated with the charge order, as observed in many ionic materials, such as cuprates (La$_{2-x}$Sr$_x$CuO$_4$\,\cite{Tranquada1995Evidence}), nickelates (hole-doped La$_2$NiO$_4$\,\cite{Tranquada1994simultaneous}), iron-based superconductors (FeSe\,\cite{Chubukov2015Origin}, KFe$_{0.8}$Ag$_{1.2}$Te$_2$\,\cite{Song2019Interwined} and Ba$_{1-x}$Na$_x$Fe$_2$As$_2$\,\cite{Avci2014Magnetically}), topological semimetals (GdSb$_x$Te$_{2-x-\delta}$ \,\cite{Lei2021}), transition metal dichalchogenides (Fe$_x$NbS$_2$\,\cite{Wu2023Discovery}), and kagome magnets (FeGe\,\cite{Yin2022Discovery, Zhao2023Photoemission}).

Fifth, the rather large charge redistribution associated with the emergent excitons would typically dictate a heavy dressing by local lattice distortions, much stronger than in the scenario of Fermi surface instabilities.
Even though such strong local deformations can be masked in the ordered component by long-range quantum fluctuations\,\cite{Yuting2015Itinerancy-Enhanced}, a significant difference in local bond length can still be observed through pair distribution function analysis of diffraction, nuclear quadrupole resonance, or X-ray absorption fine structure.
Such ``contradiction'' between probes of long- and short-wavelength has in fact been observed in transition metal dichalchogenides (NbSe$_2$\,\cite{Arguello2014, Chatterjee2015, Moncton1975Study,Weber2011}, TaS$_2$\,\cite{Joshi2019, Petkov2020}, NbTe$_4$\,\cite{Petkov2023Charge}), manganites (La$_{1.2}$Sr$_{1.8}$Mn$_2$O$_7$\,\cite{Weber2010}, Pr$_{0.5}$Sr$_{0.5}$MnO$_3$\,\cite{Zafar2024Local}), cuprates (La$_{2-x}$Sr$_x$CuO$_4$\,\cite{Reznik2006}), Pt-intermetallics (UPt$_2$Si$_2$\,\cite{Petkov2023Local}, CePt$_2$Si$_2$\,\cite{Petkov2023Local}, LaPt$_2$Si$_2$\,\cite{Petkov2023Local}, LaPt$_{2-x}$Ge$_{2+x}$\,\cite{Petkov2023Local}), and high-entropy alloys (MoNbTaW\,\cite{Andrea2024Local}).

Sixth, since the formation of bare and emergent excitons is driven by  (above-eV) high-energy \textit{short-range} physics, the transition temperature, $T_{\text{CO}}$, is determined not by the instability of a unordered Fermi liquid state, but rather by the long-range ordering of the preexisting short-range correlations.
Specifically, the dominant fluctuation for the ordering is not the typical amplitude fluctuation as in Fermi surface instability, but instead the long-range phase fluctuation that preserves the short-range structure.
Particularly in metallic ordered states, the \textit{quantum} fluctuations induced by the remaining itinerant carriers are known to decay slower in distance\,\cite{Ruderman1954Inderect,Yuting2015Itinerancy-Enhanced} and thus significant at long range.
Consequently, $T_{\text{CO}}$, is controlled by the long-wavelength phase stiffness of much lower energy, instead of the effective interaction (or the ``pseudogap'' $\Delta^*$), in contrast to the similar scale to the coherent gap, $T_{\text{CO}}\sim\Delta$, in the Fermi surface instability (or any mean-field theories on top of Fermi liquid).
This characteristic scale decoupling between $T_{\text{CO}}$ and the interaction is, in fact, typical for strongly correlated physics.
For example, the prototypical cuprates magnetically orders \textit{only} at $T_\mathrm{N}\sim 15$\,meV, in spite of the Rydberg scale formation of local magnetic moments, 150 meV scale of the emergent short-range magnetic interaction, and eV scale energy density differences between various long-range ordered states.

Seventh, the much higher energy of exciton formation immediately implies a persisting short-range correlations and lattice deformations even way above $T_{\text{CO}}$, as frequently observed in transition metal dichalchogenides (NbSe$_2$\,\cite{Arguello2014}) and cuprates (La$_{2-x}$Sr$_x$CuO$_4$\,\cite{Reznik2006}).
Correspondingly, the loss of long-range order in this scenario is associated with the melting of the exciton crystal (phase fluctuations) accompanied by exciton density fluctuations, in great contrast to the rapid vanishing of particle-hole pairs with the Fermi surface instability.
The finite-temperature phase transition of the former can thus often display a first-order nature, possibly with finite phonon frequencies, while it is typically second-order in the latter, with phonon softening to zero frequency.
This naturally explains the observation in kagome metals (CsV$_3$Sb$_5$\,\cite{Subires2023Order-disorder, Park2023Condensation, Ratcliff2021Coherent}, ScV$_6$Sn$_6$\,\cite{Lee2024Nature}, FeGe\,\cite{Zhao2023Photoemission}), iridium alloys (IrTe$_2$\,\cite{Kim2015Origin}, Lu$_5$Ir$_4$Si$_{10}$\,\cite{Becker1999Strongly}, Lu$_2$Ir$_3$Si$_5$\,\cite{Sangeetha2015Multiple}), and transition metal dichalchogenides (NbSe$_2$\,\cite{Moncton1975Study}, TaSe$_2$\,\cite{Moncton1975Study}).

Finally, concerning the quantum electronic structure, recall that within the renormalization group framework\,\cite{Wilson1971Renormalization, Nigel1992} of phases and their transitions, a system can only flow to \textit{one} particular stable fixed point that captures all the essential characteristics of the phase.
As summarized in Tab.\,\ref{tab1}, the characteristics of this scenario are \textit{qualitatively} distinct from those of the Fermi surface instability scenario (via either Fermi surface nesting or electron-phonon coupling).
It thus must belong to a fixed point distinct from the Fermi liquid.

This is apparent from its \textit{strongly correlated} structure of low-energy states, $\ket{\Psi_I; N_b}$, in Eq.\,\ref{CO}, containing in various (ordered or disordered) \textit{locations} $N_b$ emergent excitons, $\tilde{b}^\dagger_j$.(Naturally, the \textit{locations} $\{j\}_I$ are better ordered below the scale of $T_\mathrm{CO}$ but not above.)
Importantly, as ``eigen-particles''\,\cite{Hegg_Bose_metal} $\tilde{b}^\dagger_j$ \textit{fully} absorb the quantum \textit{many-body} fluctuations from other excitons, electrons and lattice distortions, particularly the strong inter-excitonic \textit{exclusive correlations}.
Consequently, $\ket{\Psi_I; N_b}$ is \textit{drastically distinct} from any product state of bare particles and the remaining itinerant carriers in the system are heavily scattered by the excitons.

Such quantum states are ``adiabatically disconnected'' from the \textit{weakly correlated} Fermi liquid,
\begin{equation}
        \ket{\mathrm{CDW}_q}=\ket{\widetilde{\mathrm{FL}}_q}\sim\prod_{k<k_\mathrm{F}}\tilde{c}^\dagger_k(q)\ket{0}\sim\prod_{k<k_\mathrm{F}}(u_k c^\dagger_k+v_k c^\dagger_{k+q})\ket{0},
    \label{CDW}
\end{equation}
obtained from the general scheme of Fermi surface instability [even with quantitatively shorter coherent length absorbed in the dressed electrons, $\tilde{c}^\dagger_k(q)$].
(Here $u_k$ and $v_k$ are coefficients.)
Therefore, with high enough density of Frenkel excitons, especially when their ordering is destroyed above $T_{\text{CO}}$, the system may easily develop non-Fermi liquid behavior, as observed in organic layered metals ($\alpha$-(BEDT-TTF)$_2$MHg(SCN)$_4$\,\cite{Merino2006NonFermiLiquid}), cuprates (YBa$_2$Cu$_3$O$_{7-\delta}$\,\cite{Arpaia2019Dynamical, Seibold2021StrangeMetal}, Nd$_{1+x}$Ba$_{2-x}$Cu$_3$O$_{7-\delta}$\,\cite{Arpaia2019Dynamical}), nickelates (La$_{2-x}$Sr$_x$NiO$_4$\,\cite{Anissimova2014Dynamic}), and kagome metals (KV$_3$Sb$_5$\,\cite{Mozaffari2024Universal}, RbV$_3$Sb$_5$\,\cite{Mozaffari2024Universal}, CsV$_3$Sb$_5$\,\cite{Mozaffari2024Universal}, ScV$_6$Sn$_6$\,\cite{Mozaffari2024Universal}).


\subsubsection*{Resolving the puzzling ordering wavevector in ScV$_6$Sn$_6$}

The scenario of Frenkel exciton crystallization offers a natural resolution to the puzzling periodicity of the observed charge order in the bilayer kagome ScV$_6$Sn$_6$, another member of the same V-based kagome superconductor family as CsV$_3$Sb$_5$ discussed above.
On the one hand, similar to the latter, nearly all observed physical properties\,\cite{Arachchige2022Charge} of ScV$_6$Sn$_6$ coincide with the above generic characteristics.
On the other, while the nearly nested Fermi surface of ScV$_6$Sn$_6$ resembles very much that of CsV$_3$Sb$_5$, the observed\,\cite{Korshunov2023Softening} in-plane charge ordering wavevector, $\sqrt{3}\times \sqrt{3}$, is drastically distinct from the more nesting-compatible $2\times 2$ wavevector observed in CsV$_3$Sb$_5$.

Given the above-mentioned strong local lattice distortion associated with their formation, naturally the Frenkel excitons in the bilayer structure of ScV$_6$Sn$_6$ have a strong tendency toward interlayer binding for an energetically more compatible structure between layers.
Such a bound interlayer ``bi-exciton'' would have an enhanced effective binding (lower $E^\mathrm{F}$) and a reduced kinetic strength $t$ in Eq.\,(\ref{eq2}).
As demonstrated in Fig.\,\ref{fig4}(d), both of these effects would favor a smaller exciton size, from $2\times 2$ (blue) to $\sqrt{3}\times \sqrt{3}$ (red), thus offering a qualitative explanation for the puzzling nesting-incompatible ordering wavevector.

The proposed exciton crystallization also intuitively explains the other puzzle on the variable out-of-plane period of the observed charge correlation, namely a $\sqrt{3}\times \sqrt{3}\times 2$ short-range order coexisting with a $\sqrt{3}\times \sqrt{3} \times 3$ long-range order\,\cite{Arachchige2022Charge, Cao2023Competing, Korshunov2023Softening, Pokharel2023Frustrated, Wang2024origin}.
Similar to the case of CsV$_3$Sb$_5$, the out-of-plane stacking of the bi-exciton layers simply follows a lower-energy closed packing.
Specifically, as shown in Fig.\,\ref{fig5}(c)(d), in contrast to the four allowed sites for stacking in the $2\times 2$ excitonic lattice in panel (a)(b), the $\sqrt{3}\times \sqrt{3}$ excitonic lattice has three allowed sites.
In turn, the chiral ABC stacking, consistent with the observed $\sqrt{3}\times \sqrt{3} \times 3$ period, would have the lowest-energy.
As in the case of CsV$_3$Sb$_5$, slightly higher-energy ABAB stacking is also locally stable, thus the presence of the observed short-range order.

In summary, to address the long-standing issue of unconventional charge order in functional materials, we propose an alternative general scenario - crystallization of long-lived Frenkel excitons - applicable to ionic systems hosting strong short-range correlation.
We demonstrate this mechanism via the recently discovered kagome superconductors, CsV$_3$Sb$_5$, which naturally offers an intuitive explanation for the distinct charge-order patterns in ScV$_6$Sn$_6$.
Furthermore, we compare the generic characteristics of this mechanism against those of the standard picture of Fermi surface instability, which indicates a broad applicability of this mechanism as a unifying paradigm for understanding diverse charge-ordered phenomena in a wide range of ionic systems.
The proposed framework thus offers a long-sought theoretical complement to modern descriptions on unconventional charge order in correlated functional materials.

\section*{Acknowledgements}
We thank the helpful discussion with Chi Ming Yim and Yu Zheng about the STM measurements. This work is supported by a UKRI Future Leaders Fellowship [UKRI2083] and from an EPSRC grant [EP/V062654/1]; by the Gianna Angelopoulos Programme for Science, Technology, and Innovation; by the National Natural Science Foundation of China (NSFC) \#11674220 and \#12042507; by the Innovation Program for Quantum Science and Technology 2021ZD0301900; and by the Shanghai Branch, Hefei National Laboratory, Shanghai 201315, China.

\bibliography{main.bib}

\end{document}